\documentclass[fleqn,usenatbib,usedcolumn]{mnras}

\usepackage{amsmath}	
\usepackage{amssymb}	
\usepackage{threeparttable}
\usepackage[justification=centering]{caption}
\usepackage{aecompl}
\usepackage{float}

\usepackage[british]{babel}             
\usepackage{txfonts}                    
\let\la=\lesssim 
\usepackage[T1]{fontenc}                
\usepackage{graphicx}                   

\title[A possible seismic signature of heavy elements ionization]{The CoRoT target HD49933: a possible seismic signature of heavy elements ionization in the deep convective zone.}

\author[Brito \& Lopes]{
Ana Brito,$^{1,2}$\thanks{E-mail: ana.brito@tecnico.ulisboa.pt}
Il\'idio Lopes,$^{1}$\thanks{E-mail: ilidio.lopes@tecnico.ulisboa.pt}
\\
$^{1}$Centro Multidisciplinar de Astrof\'{\i}sica, Instituto Superior T\'ecnico, Universidade Tecnica de Lisboa \\ Av. Rovisco Pais, 1049-001 Lisboa, Portugal \\
$^{2}$Departamento de Matem\'atica, Instituto Superior de Gest\~ao, Lisboa - Portugal\\
}

\date{Accepted XXX. Received YYY; in original form ZZZ}

\pubyear{2015}

\begin{document}
\label{firstpage}
\pagerange{\pageref{firstpage}--\pageref{lastpage}}
\maketitle

\begin{abstract}
	We use a seismic diagnostic, based on the derivative of the phase shift of  the reflected by the surface acoustic waves, to probe the outer layers of the star HD 49933. This diagnostic is particularly sensitive to partial ionization processes occurring above the base of the convective zone. The regions of partial ionization of light elements, hydrogen and helium, have well known seismological signatures. In this work we detect a different seismic signature in the acoustic frequencies, that we showed to correspond to the location where the partial ionization of heavy elements occurs. The location of the corresponding acoustic glitch lies between the region of the second ionization of helium and the base of the convective zone, approximately 5\% below the surface of the stars. 
\end{abstract}






\section{Introduction}


\begin{table*}
	\large
	\begin{threeparttable}
		\caption{Parameters of the optimal models obtained for the CoRoT target HD 49933 with the code CESAM.}
		\begin{tabular}{ c c c c c c c c c c  }          
			\hline
			\hline
			Model &${M}/{M}_\odot$ & ${R}/{R}_\odot$      & ${L}/{L}_\odot$     & $T_{\text{eff}}(K)$      & Age (Gyr)    & $(Z/X)_s$   & $\alpha$ & log $g$ & $r_{\text{bcz}}/R$\\
			\hline
			\hline
			1 & 1.15  &1.42 & 3.35 & 6561  & 3.152 & 0.00560 &  1.7 & 4.19 & 0.874  \\
			2 & 1.25  &1.45 & 4.21 & 6800  & 1.981 & 0.00549 &  1.8 & 4.21 & 0.930  \\
			3 & 1.20  &1.43 & 3.76 & 6728  & 2.582 & 0.00550 &  1.9 & 4.20 & 0.890  \\
			4 & 1.15  &1.44 & 3.67 & 6676  & 3.746 & 0.00822 &  2.3 & 4.18 & 0.802  \\  
			\hline
			\hline
		\end{tabular}
		\label{table:1}
	\end{threeparttable}
\end{table*}

\begin{figure}
	\includegraphics[width=0.5\textwidth]{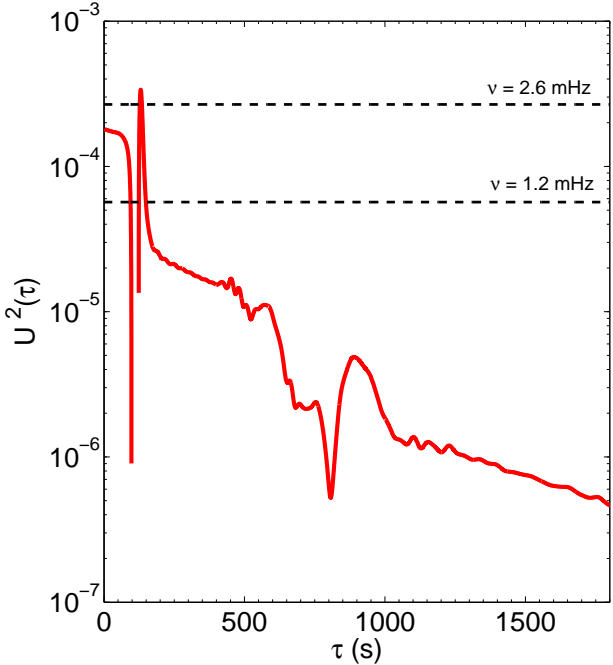}
	\caption{Reflecting acoustic potential for model 1: a theoretical qualitative description of the outer envelope of the star HD 49933. This acoustic potential was computed between the surface of the star and the base of the convective zone (BCZ). Clearly visible in this figure are, the superadiabatic region and the region of ionization of light elements. Horizontal dashed black lines represent the frequencies of modes with 1.2 mHz and 2.6 mHz. This range of frequencies correspond to the observational window.}
	\label{fig: 1}
\end{figure}

\begin{figure}
	\includegraphics[width=0.5\textwidth]{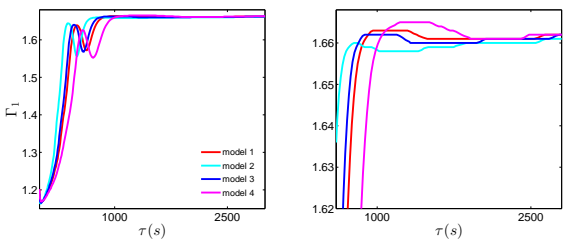}
	\caption{Left: Profiles of the first adiabatic exponent, $\Gamma_1$, for all the models as a function of the acoustic depth $\tau$. The depression of $\Gamma_1$ around 500--900 s is due to the ionization of light elements. Right: Slight depression of $\Gamma_1$ due to the ionization of heavy elements.}
	\label{fig: 2}
\end{figure}

The main sequence star HD 49933, a solar-type star of spectral class F5, has been observed since a few years ago. First from the ground \citep{2005A&A...431L..13M}, and then as a primary target of the CoRoT mission \citep{2006ESASP1306...33B, 2008Sci...322..558M}. Several authors obtained with different methods their fundamental parameters ($T_{\text{eff}}, [\text{Fe/H}], \text{log g}$). These studies point to a metal poor star, slightly hotter than the Sun \citep{2009A&A...506..235B, 2009A&A...506..203R, 2010A&A...510A.106K, 2015A&A...582A..49H, 2015A&A...582A..81J}. The acoustic oscillations exhibited by the star are thought to be driven stochastically by convective motions in the subsurface layers. However, the seismic analysis of this star was not straightforward. The \'echelle diagram \citep{1983SoPh...82...55G} has been widely used in asteroseismology by observers to organize and display the mode frequencies of solar like oscillations. The frequency ($\nu$) is plotted versus the  modulo of the large separation ($\nu \; \text{mod} \; \Delta \nu$). This kind of representation originates a stack of frequencies aligned vertically by mode degree. Consequently, irregularities are easily detectable and manifest themselves as a curvature of the vertical ridges of the \'echelle diagram, which in turn, are known to be related to surface effects and acoustic glitches. As an F-type star, the star HD 49933 is an emblematic case of the "F star problem", which consists in a difficulty in the identification of the modes due to their short lifetimes, and consequently, to their large line widths. This situation was observed for the first time, precisely, in this star. Indeed, the mode line widths increase with increasing effective temperature \citep[e.g.,][]{2012ApJ...757..190C}. In some hot stars the mode line width become so large that makes it difficult to distinguish between the ridges $l=0,2$ and $l=1$ in the \'echelle diagram, thus originating two possible scenarios of mode identification. \citet{2008A&A...488..705A} was the first who tried to resolve this ambiguity by fitting the power spectrum for both possible scenarios of identification. He then favored one of the scenarios, known in literature as scenario A. Further studies followed without a definite and conclusive mode identification \citep{2009A&A...506...15B, 2009A&A...506.1043G, 2009A&A...506..435R, 2010A&A...510A.106K}. However, with additional time of observation and improved methods, \citet{2009A&A...507L..13B} reverted to the initial mode identification. This second scenario of mode identification is known as scenario B. The work presented in this paper is based on the mode identification made by \citet{2009A&A...507L..13B}.
\begin{figure*}
	\includegraphics[width=0.3\textwidth]{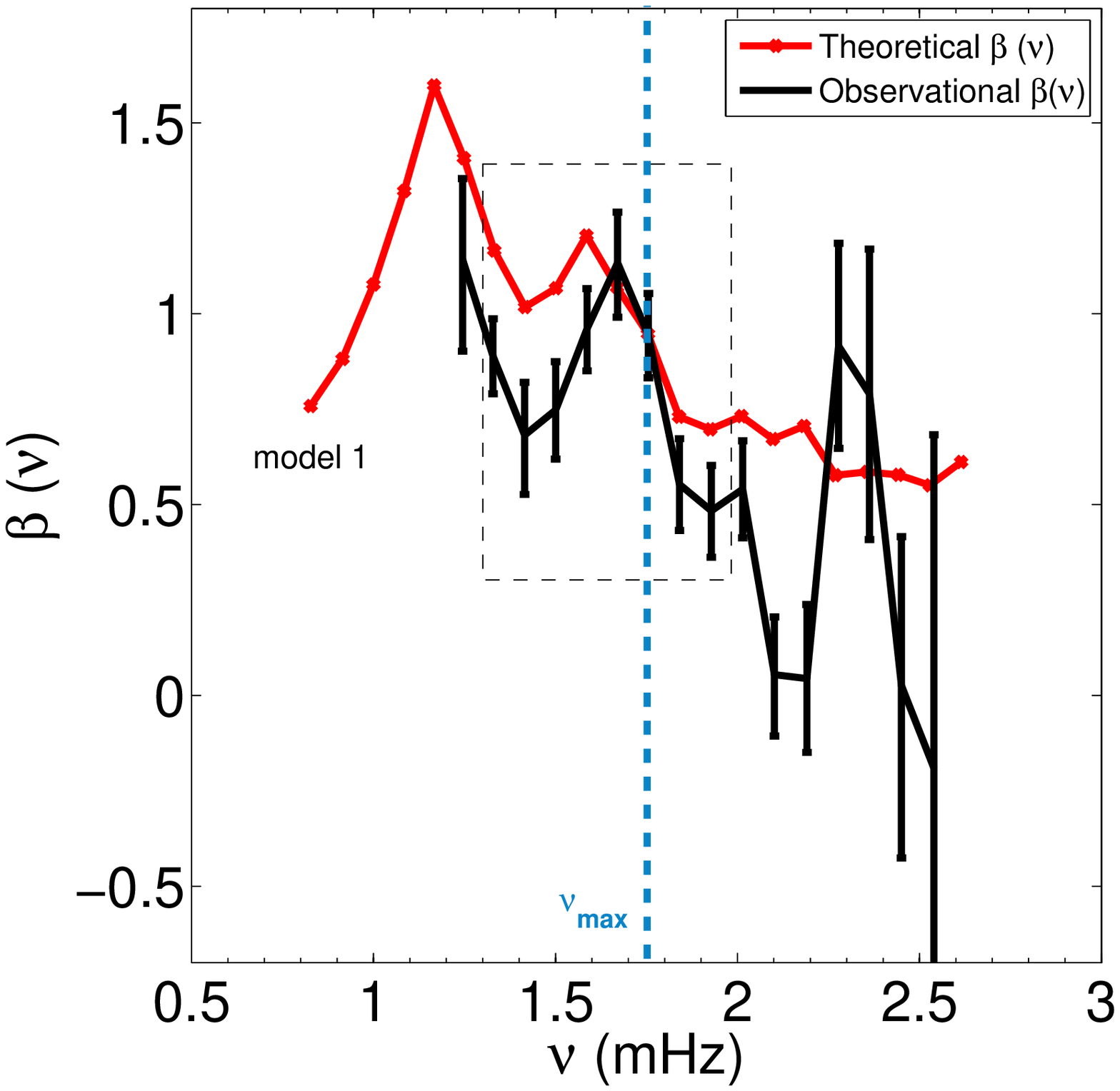}
	\includegraphics[width=0.3\textwidth]{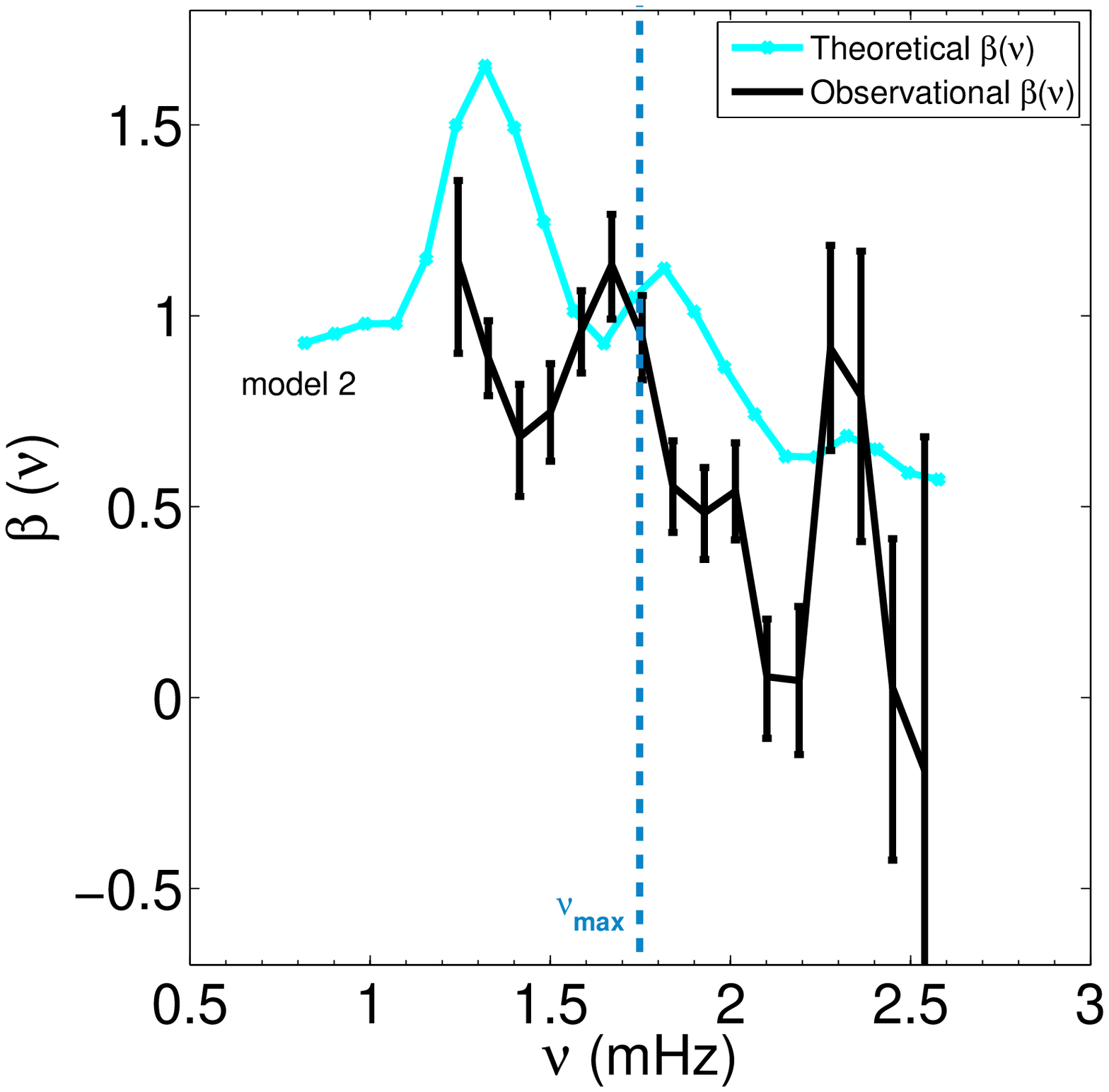}\\
	\includegraphics[width=0.3\textwidth]{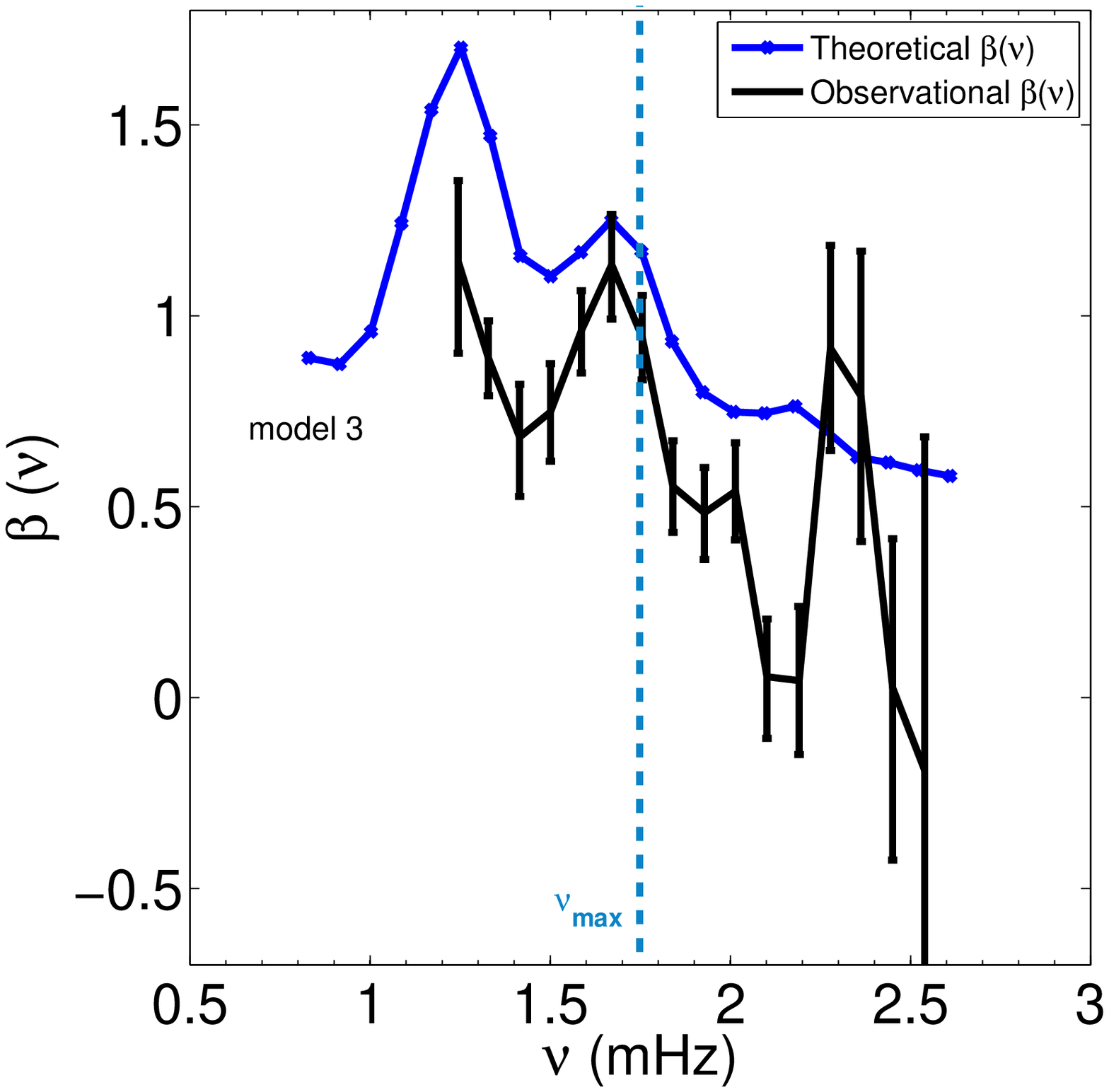}
	\includegraphics[width=0.3\textwidth]{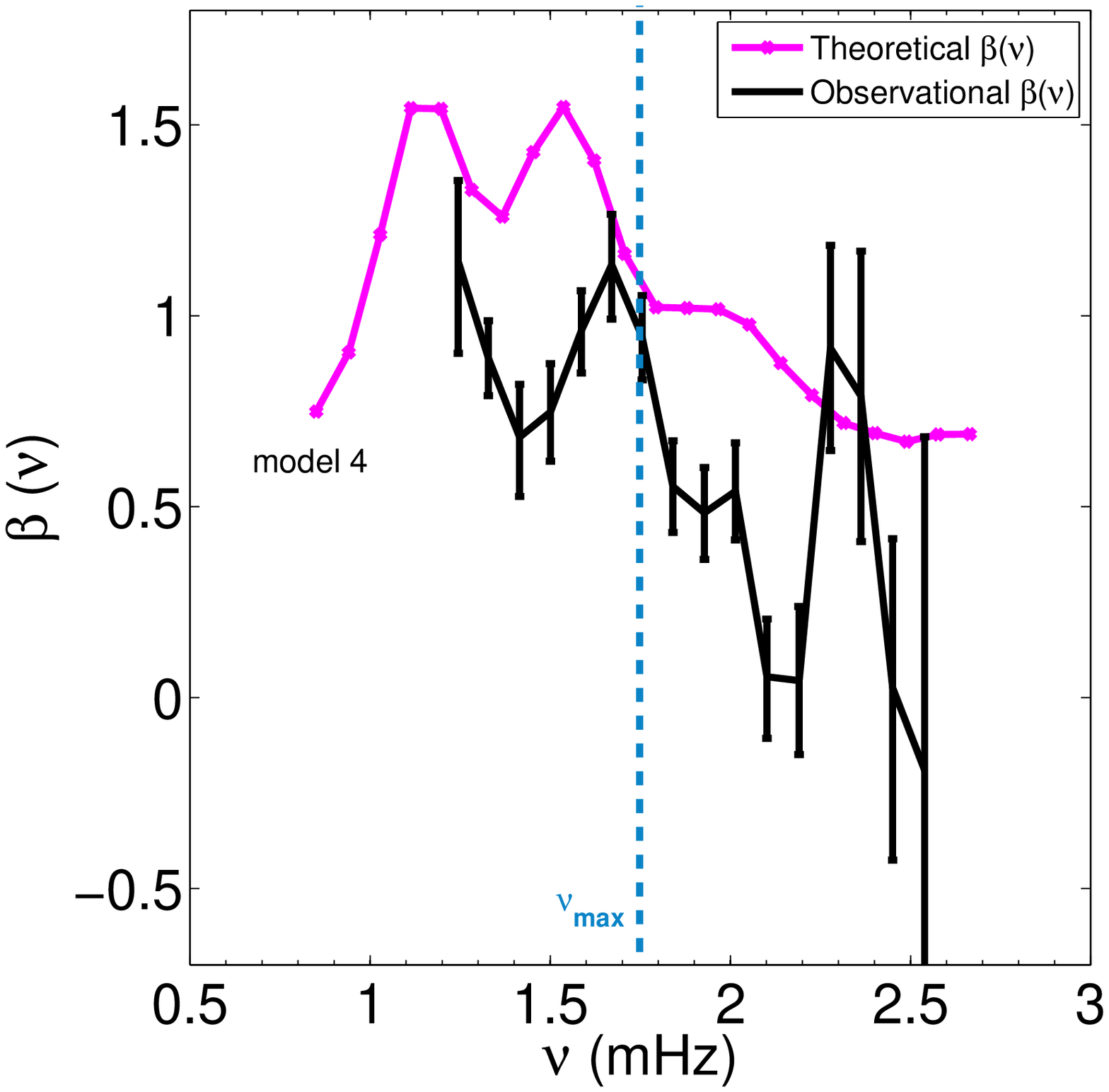}
		\caption{Comparison of the theoretical seismic parameter with the observable $\beta(\nu)$. Black line represents the seismic parameter computed from the observational frequency table \citep{2009A&A...507L..13B} with the corresponding error bars. Each model is represented by the respective color and the $\beta(\nu)$ was computed from theoretical tables of frequencies obtained with the code ADIPLS \citep{2008Ap&SS.316..113C}. Dashed vertical blue line shows the location of the frequency of the maximum power around which the best agreement is achieved.}
	\label{fig: 3}
\end{figure*}
In regards to magnetic activity, the CoRoT target HD 49933 was the first star for which asteroseismology could claim the discovery of a magnetic cycle. \citet{2010Sci...329.1032G} were the first to observe variations with the activity in the p-mode frequencies and amplitudes of the star. These variations were confirmed recently by \citet{2016A&A...589A.103R} and present a pattern of anti-correlation between frequencies and amplitudes in the same way they are observed in the Sun. They yield a magnetic cycle with a period of 120 days. The frequency shifts increase with increasing frequency and reach a maximum of $2 \, \mu$Hz in the vicinity of $2100 \, \mu$Hz \citep{2011A&A...530A.127S}. This short magnetic cycle is confirmed by spectroscopy with a Mount Wilson S-index value of 0.3 \citep{2010Sci...329.1032G}. 

Nowadays, the outer layers of solar-type stars are still far to be well described by theoretical models. This is because, in these layers, there are different and complex physical processes taking place in the same timescales. There are turbulent motions of the external convective zone and convective overshooting;  partial ionization regions of the most abundant chemical elements with the associated abrupt changes in the properties of stellar matter; driving and damping of oscillation modes; effects of non-adiabaticity; differential rotation and magnetic fields generation through the dynamo mechanism.
Asteroseismology, as it is well known, is a powerful tool to infer the internal structure of stars using the information stored in the oscillation frequencies of the acoustic modes. Yet, in the outer layers, the intricate web of physical processes mentioned above, will manifest itself in the frequencies of the p-modes. All current theoretical models fail in modeling the outer stellar layers. There is a systematic difference between the theoretical frequencies and the observational frequencies, even in the solar case. The most affected modes by the outer layers are the high frequency modes since the upper turning points of high frequency waves are located closer to the surface of the star. Therefore, the systematic offset between theory and observation increases with increasing frequency. \citet{2008ApJ...683L.175K} suggested a power law to correct this systematic difference. Nevertheless, when studying specifically this external region of the star, the correction may blur or distort important information contained in this part of the spectrum. 

The phase shift experienced by the sound waves when they are reflected by the surface of the star, is a tracer that can be used to extract information about these more external layers. Specifically, in this paper we use the derivative of the phase shift, the so-called seismic parameter $\beta(\nu)$, to extract information about the outer layers of the CoRoT target HD 49933. It is known that the oscillation pattern, i.e., phase and amplitude, of the seismic signature $\beta(\nu)$, allows to characterize the partial ionization zones of the Sun and solar-type stars \citep{1997ApJ...480..794L, 2001MNRAS.322..473L}. In the case of the Sun it allowed the inference of the solar photospheric abundance of helium \citep{1991Natur.349...49V}.  By direct comparison of the theoretical and observational signatures of $\beta(\nu)$ we aim to infer about the ionization processes taking place in the outer layers of HD 49933.

We computed four theoretical models, fully compatible to the other models of this star described in literature \citep{2009A&A...506..235B, 2011A&A...534L...3B, 2009CoAst.158..285O, 2009A&A...506..175P, 2010A&A...510A.106K, 2011JPhCS.271a2038C, 2014RAA....14..683L, 2015A&A...574A..45R}. Our goal is to probe, from a seismological point of view, the outer layers of the star HD 49933. In section \ref{sec:2} we describe and apply the seismic diagnostic $\beta(\nu)$ to obtain a detailed characterization of the outer layers of the star.
Section \ref{sec:3} provides a description of the partial ionization profiles of chemical elements in the outer layers of the star, with special emphasis on the role of the heavy elements. In section \ref{sec:4}, we analyze the oscillatory component of the observed frequencies and relate it with a precise location in the convective zone of the star. Finally, in section \ref{sec:5} we present some conclusions about the relationship between ionization processes in the outer layers of the star and oscillatory characteristics of the observed and theoretical seismic parameter $\beta(\nu)$.


\section{The structure of the acoustic cavity of the star HD 49933}
\label{sec:2}

\begin{figure*}
	\includegraphics[width=0.9\textwidth]{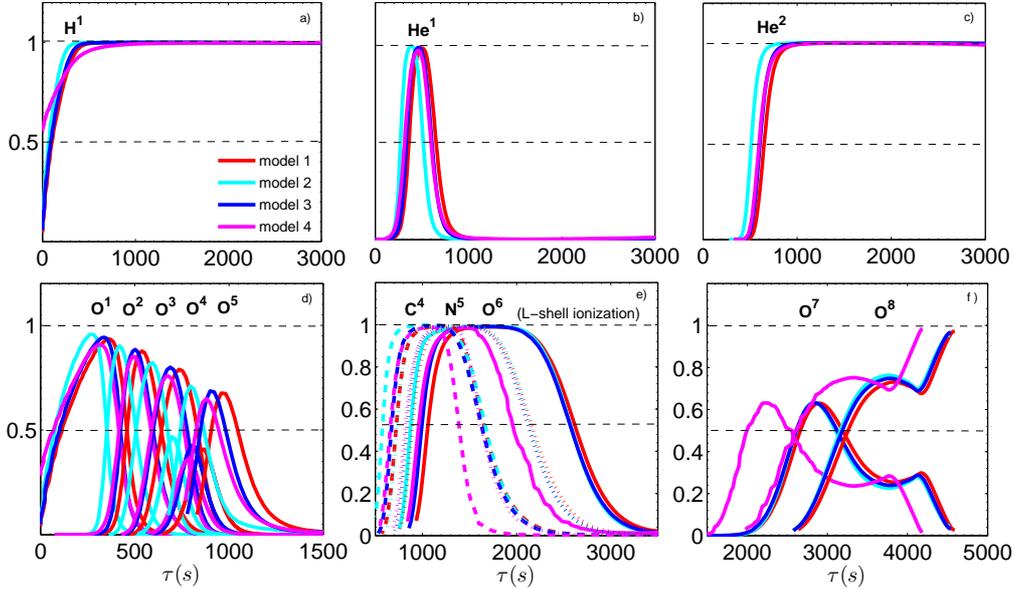}
		\caption[LoF entry]{{\bf{Upper panel}}: degrees of ionization for hydrogen and helium, of all the models, plotted against the acoustic depth $\tau$. Light elements are almost completely ionized at an acoustic depth of 900 s. a) Ionization of hydrogen. b) First ionization of helium. c) Second ionization of helium. {\bf{Bottom panel}}: degrees of ionization for heavy elements of all the models, plotted against the acoustic depth $\tau$. In d) it is shown the ionization of the first five valence electrons for oxygen. Carbon and nitrogen have similar patterns of ionization for the first 3 and 4 valence electrons respectively, and to avoid overloading the figure these are not represented. e) shows the ionization of the last electron of the L-shell for carbon (dashed lines), nitrogen (dotted lines) and oxigen (solid lines). The ionization of the last electron of the L-shell, for CNO, occurs in the adiabatically stratified part of the convective zone, in a region where hydrogen and helium are already almost fully ionized. f) represents the ionization of elements of the K-shell for oxygen. K-electron ionizations, although very interesting, occur in the deep interior of the star and are behind the scope of this work. Here again we have plotted only the oxygen in order not overload the figure. 
		Horizontal black dashed lines represent the levels of full ionization and where the element is half ionized for the corresponding degree of ionization.}
	\label{fig: 4}
\end{figure*}

\subsection{The phase-shift diagnostic of the external layers}

The complete fourth order set of differential equations describing the adiabatic stellar oscillations, can be reduced, in the Cowling approximation, to a second order Schr\"{o}dinger type equation \citep [e.g.][]{1989ASPRv...7....1V, 2001MNRAS.322..473L}

\begin{equation} \label{eq:(1)}
	\frac{d^2 \psi}{d\tau^2} + (\omega^2 - U^2)\psi = 0\,,
\end{equation} 
where 
\begin{equation} \label{eq:(2)}
	\tau = \int_r^R \frac{d r}{c} 
\end{equation}
is the acoustic depth, and $U^2(\tau)$ is the reflecting acoustic potential, which is determined by the structure of the envelope of the star. This straightforward outcome is achieved using the acoustic depth $\tau$ as an independent variable and $\psi = r(\rho c)^{\frac{1}{2}} \xi$ as a function of the radial displacement $\xi$ \citep{1979nos..book.....U}.
The reflecting acoustic potential is given by	
\begin{equation} \label{eq:(3)}
	U^2(\tau) = \frac{g}{c} \left( \frac{g}{c} - \frac{d\ln h}{d\tau} \right) + \left[ \frac{1}{2} \frac{d \ln \zeta}{ d \tau} \right]^2 - 		\frac{1}{2}\frac{d^2 \ln \zeta }{d\tau^2}
\end{equation}
where
\begin{equation} \label{eq:(6)}
	\zeta = \frac{r^2 h}{c}
\end{equation}
and 
\begin{equation} \label{eq:(7)}
	h(r) = \rho^{-1} \exp \left( -2 \int_0^r \frac{g}{c^2} dr \right) .
\end{equation}
In the above $g$ is the gravitational acceleration, $c$ and $\rho$ are the sound speed and the density respectively, and $r$ the stellar radius. 
Below the outermost layers, in the region ${U^2\ll \omega^2}$, the solution of equation (\ref{eq:(1)}) can be expressed in the form
\begin{equation}
	\psi \approx A \cos\left[\omega \tau - \frac{\pi}{4} - \pi \alpha (\omega) \right],
\end{equation}
where $\alpha(\omega)$ is the outer phase shift of the sound wave reflected from the potential $U^2(\tau)$, with $A$ being the amplitude of the wave. Moreover, $\omega = 2\pi \nu$ is the frequency and $\nu$ is the cyclic frequency.
It is well known that, in the context of the asymptotic theory of stellar oscillations, there is a relation between the outer phase shift $\alpha(\omega)$ and the structure of the stellar envelope \citep[e.g.][]{1994MNRAS.268..880R, 1997ApJ...480..794L}. For each stellar model the dependence of the outer phase shift on the frequency can be computed from the corresponding acoustic potential \citep{1997ApJ...480..794L}.

In the outer layers of a solar type star the sound speed is low. Consequently, for low degree modes of oscillation, the vertical component of the wave vector is much greater than the horizontal. This means that the phase shift is a function of the frequency alone $\alpha = \alpha(\omega)$. The outer phase shift will carry all the uncertainties of these more superficial layers and also the features associated with non adiabatic processes taking place in these regions, and thus, it can be used as a seismic probe of the outermost layers of a solar type star.
The information contained in the dependence of the phase shift on the mode frequencies can be obtained from the derivative of the phase shift
\begin{equation}
	\beta(\omega) =\beta(\nu)= - \omega^2  \frac{d}{d \omega} \left( \frac{\alpha}{\omega} \right) ,
\end{equation}
according to the technique described by e.g. \citet{1987SvAL...13..179B, 1989SvAL...15...27B, 1989ASPRv...7....1V}.
This function is known as the seismic parameter $\beta(\nu)$. It allows to obtain information about the phase shift of the reflected sound waves from a model envelope and from a table of observational/theoretical frequencies. Hence, we can explore the effect of the outer layers of the star on the oscillation spectrum.
				
\subsection{The reflecting acoustic potential of the star HD 49933}
				
The star HD 49933 has been modelled a few times by different authors \citep[e.g.][]{ 2009A&A...506..175P, 2009CoAst.158..285O, 2011A&A...534L...3B, 2011JPhCS.271a2038C, 2014RAA....14..683L, 2015A&A...574A..45R}. We use the stellar evolution code CESAM 2k to generate a grid of models with different values for the mass, metallicity and mixing length parameter. Our grid of models reproduce the observed values of $T_{\text{eff}}$, log $g$, $(Z/X)_s$ and $\Delta \nu$, within the observational errors \citep{2009A&A...506..235B, 2009A&A...506..203R, 2015A&A...582A..49H, 2015A&A...582A..81J}.
The models use nuclear reaction rates computed with the NACRE compilation \citep{1999NuPhA.656....3A}, the solar mixtures of \citet{2009ARA&A..47..481A}, OPAL 2001 as equation of state and the most recent OPAL opacities. The effect of diffusion was included using the Burgers formalism and the atmosphere is given by the Eddington law. The contact of the atmosphere with the upper part of the envelope is at the optical depth of 20. Mixing-length theory \citep{1958ZA.....46..108B} was used to describe convection.
Finally, using a $\chi^2$ minimization to reproduce the observational constraints as closely as possible we choose four best models. The main characteristics of the models are listed in Table \ref{table:1} and the results reflect the diversity of models in the literature. 
				
Figure \ref{fig: 1} represents the acoustic potential computed for model 1 of the star HD 49933. This acoustic potential is
represented between the surface of the star, which we consider in this work to be the location of the temperature minimum ($\tau = 0$ s), and the base of the convective zone. The acoustic potential for the  solar case was object of detailed and numerous studies \citep[e.g][]{1987SvAL...13..179B, 1989SvAL...15...27B, 2000MNRAS.316...71B, 2001MNRAS.322..473L}. 

The first adiabatic exponent, $\Gamma_1 = \left(\frac{\partial \ln p}{\partial \ln \rho} \right)_s$ is determined by the equation of state, and below the superadiabatic region follows an adiabatic stratification. The ionization of all chemical elements -- light and heavy -- causes a lowering of the first adiabatic exponent $\Gamma_1$. The magnitude of the lowering should depend on the abundances of the elements as noted by \citet{2000MNRAS.316...71B}. The $\Gamma_1$ profile for the set of models is shown in Figure \ref{fig: 2}. The decrease of $\Gamma_1$ due to ionization of light elements is located in the region where the structure of the convective envelope obeys a stratification close to adiabatic. This depression has a well known effect in the behavior of the sound speed, which in turn affects the oscillation modes. For this reason $\Gamma_1$ plays an important role in seismological analysis. Less studied but very interesting is the depression of $\Gamma_1$ due to ionization of heavy elements  \citep{2011JPhCS.271a2036B}. 
In stars slightly more massive than the Sun, the pressure is reduced in the ionization zones and this produces an effect in the adiabatic exponent $\Gamma_1$, which in turn has an explicit relation with the acoustic potential \citep{1991SoPh..133..149M}. 
					
\begin{figure}
	\centering
	\includegraphics[width=0.4\textwidth]{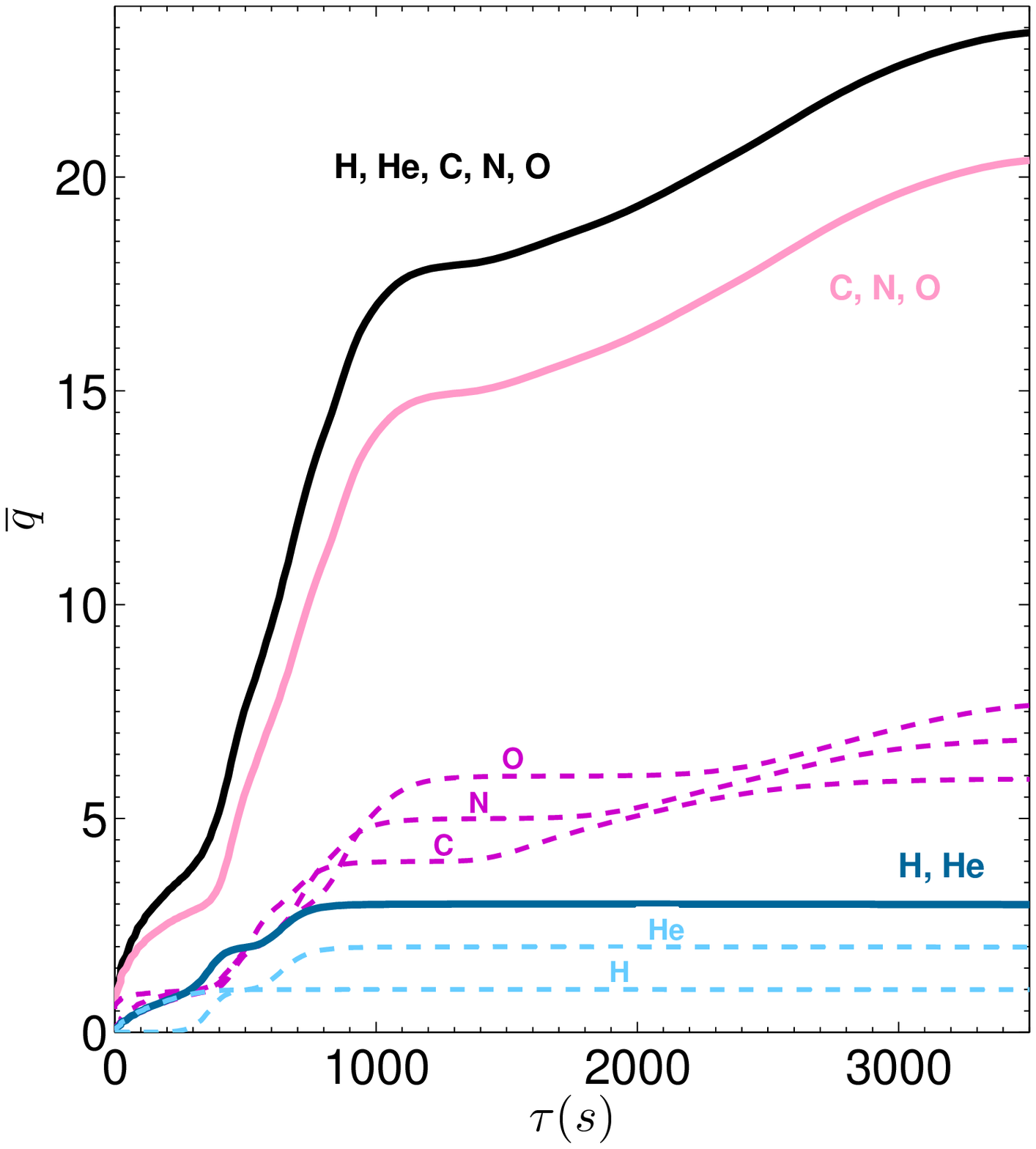}
	\includegraphics[width=0.4\textwidth]{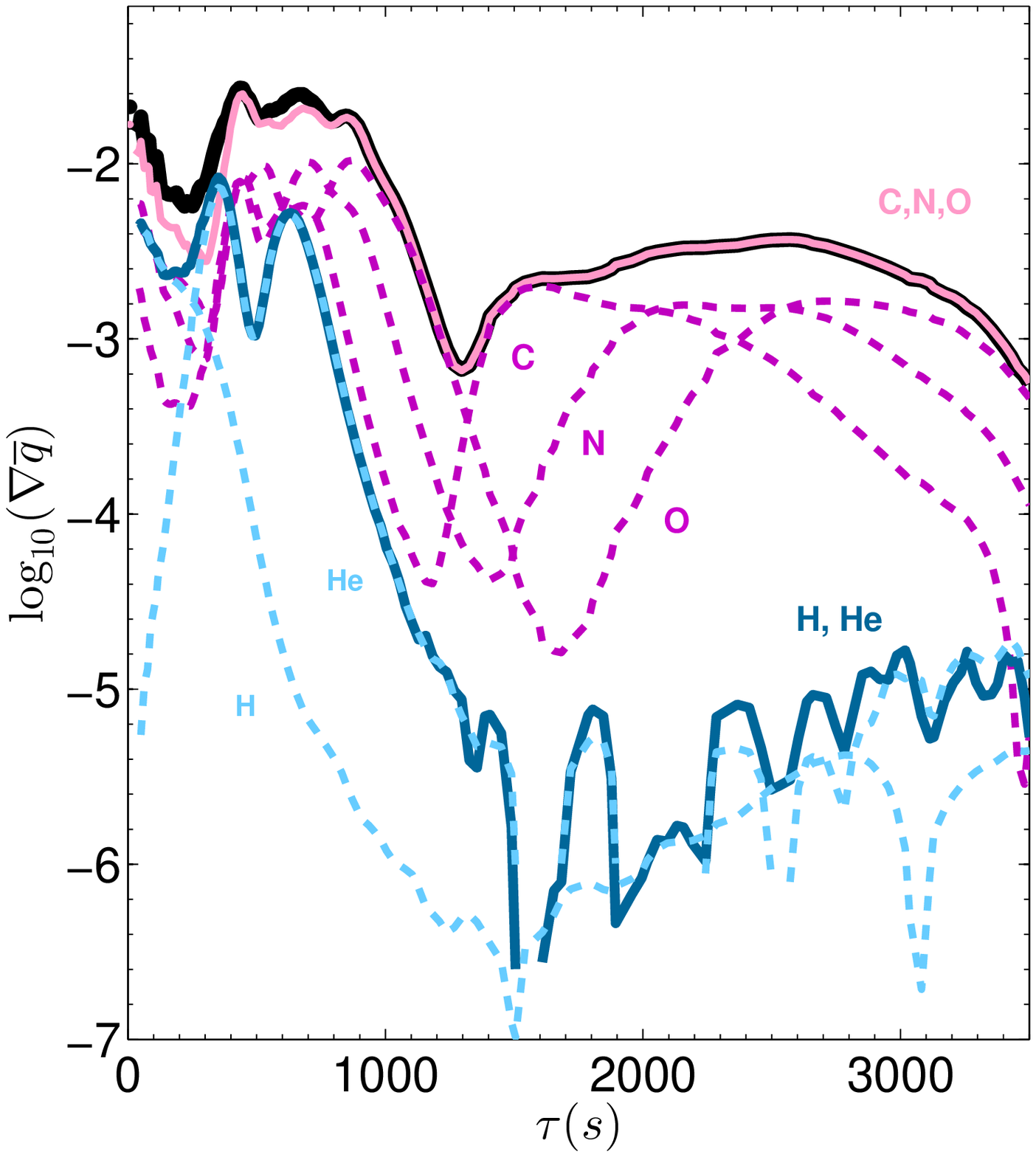}
	\caption{{\bf{Top:}} Average ionic charges of chemical elements. Dashed blue lines: hydrogen and helium. Dashed purple lines: carbon, nitrogen and oxygen. Solid black like represents the effective mean ionic charge as defined by equation (\ref{eq: 1}). Solid pink line is the contribution of heavy elements whereas solid gray blue line is the contribution of light elements. {\bf{Bottom:}} Gradient of effective mean ionic charge and the individual contributions of the light (hydrogen and helium) and heavy elements (carbon, nitrogen and oxygen) to this total gradient. All curves were represented only for model 1 since all the other models have similar contributions.}
	\label{fig: 5}
\end{figure}
				
\subsection{Recovery of $\beta(\nu)$ from theoretical and observational frequencies}

\begin{figure*}
	\includegraphics[width=0.9\textwidth]{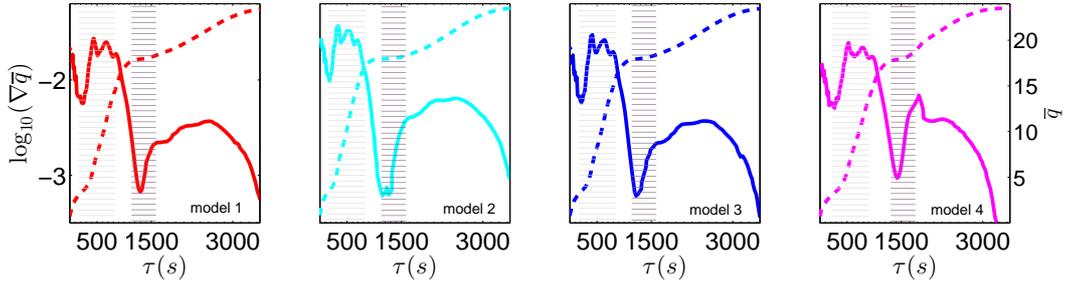}
	\caption{The effective mean ionic charge (dashed line) and the correspondent gradient (solid line) versus the acoustic depth $\tau$. The abrupt variations of the gradient are related with the changes of slopes in the effective mean ionic charge. Gray shaded bars indicate the variations and also pinpoint two distinct regions in the stellar envelope. Region 1 is dominated by ionization of light elements and region 2 is dominated by ionization of heavy elements.}
	\label{fig: 6}
\end{figure*}	
		
The matching of asymptotic solutions in the interior part of the star, with the exact solution at the surface (here the asymptotic theory is not valid) leads to the following equation for the eigenfrequencies, a dispersion relation known as the Duvall law,					
\begin{equation}
	F(w) = \pi \frac{\alpha(\omega)}{\omega} + \pi \frac{n}{\omega},
\end{equation}
where $w=\omega/L$, $L=l+1/2$ and $n$ and $l$ are respectively, the radial order and the degree of the mode. Since $F(w)$ is determined by the sound speed in the stellar interior, this equation provides a quantitative relation between the phase shift $\alpha(\omega)$ and the structure of the outer layers of the star.
As mentioned above the direct comparison between the seismic observable $\beta(\nu)$, calculated from observational frequencies, with the $\beta(\nu)$ obtained from theoretical frequency tables allows us to study the effect of the structure of the acoustic cavity on the oscillation spectrum. We compute $\beta(\nu)$ according to
\begin{equation}
	\beta_{l,n}(\nu) = \frac{\nu_{n,l} - L\frac{\partial \nu}{\partial L} - n \frac{\partial \nu}{\partial n}}{\frac{\partial \nu}{\partial n}}.
\end{equation}
The algorithm of the derivatives is based on central differences and  the observational frequency  table was taken from \citet{2009A&A...507L..13B}. To each model we have plotted the theoretical seismic parameter $\beta(\nu)$ against the cyclic frequency $\nu$ and compare it with the observational parameter. The results are shown in Figure \ref{fig: 3}  for $l=1$ modes over a range of frequencies from 1.2 mHz to 2.6 mHz in the case of observational frequencies, and from 0.78 mHz to 2.7 mHz in the case of theoretical frequencies. The use of the observational mode $l=1$ is justified by the following. First, the acoustic modes with degree $l=1$ are the modes with the best observational accuracy, i.e., with smaller error bars. Second, since we are using central differences to compute the numerical derivatives, $l=1$ modes are the modes which less suffer from boundary effects. Besides, it is known that the success of any numerical differentiation is tightly connected with the structure of the data. In this work, the minimal distance between frequencies is around $45 \, \mu$Hz, whereas the accuracy of the data itself is $\sim 1 \, \mu$Hz. Then the errors associated with the derivatives can be considered acceptable.
				
It is known that regarding the Sun, the period of the sinusoidal component in $\beta(\nu)$ is related to the location of the second ionization of helium whereas the amplitude of this sinusoid contains information about the helium abundance in the envelope \citep{1991Natur.349...49V}. It is also clear from Figure \ref{fig: 3} that the best agreement between theory and observations is achieved around the frequency of maximum power $\nu_{\text{max}}$. In the high frequencies we have a severe disagreement between all theoretical models and observational $\beta (\nu)$. 
		

\section{Partial ionization profiles in the outer layers of the star HD 49933: the role of heavy elements}
\label{sec:3}

Microscopic processes occurring in stellar interiors are closely related with macroscopic characteristics, some of which, we can observe and measure. Thermodynamic properties of the stellar matter are thus crucial to explain and understand the macro/micro connections of the stellar structure. These thermodynamic properties depend on the degree of ionization of the stellar matter. With the increasing of the depth and consequently the increasing of temperature (and density, and pressure), elements which are practically neutral at the surface, become progressively ionized. All chemical elements undergo processes of ionization from the surface to the interior of the star. Depending on the atomic configuration of the element and on the thermodynamic properties of the star, different patterns of ionization profiles will emerge and reflect different internal structures. In Figure \ref{fig: 4} we plot the degrees of ionization for light elements (upper panel) and heavy elements (bottom panel) against the acoustic depth $\tau$. Contrarily to light elements that are fully ionized well before the base of the convective zone, the ionization of heavy elements takes place through almost the whole star.

\subsection{The gradient of the effective mean charge}

To understand the impact of the ionization processes in the outer layers of the star HD 49933, specifically the ionization of heavy elements, we compute the distribution with the depth of an \emph{effective mean ionic charge}. We first define this charge as

\begin{equation}
	\bar{q} = \sum_i \langle Q_i \rangle \, , 
	\label{eq: 1}
\end{equation}
where $\langle Q_i \rangle$ is the mean ionic charge of the element $i$. Then we take the gradient of this effective charge with respect to the acoustic depth to identify the regions with larger variations of mean ionic charge. Figure \ref{fig: 5} shows the average ionic charges for each element as well as the effective mean ionic charge $\overline{q}$. Also represented are the contributions of light elements (hydrogen, helium) and the contributions of heavy elements (carbon, nitrogen, oxygen). Figure \ref{fig: 5} also shows the distribution with the depth of the gradient of the mean charge. We can spot two distinct regions where there is a steep rise of the ionic charge, and consequently, a large variation of the gradient. These regions are indicated with vertical shaded bars in Figure \ref{fig: 6}. The first region, closer to the surface, (region 1) is clearly associated with the ionization of light elements, and depending on the model, stretches from $ 400 $ s to $1000$ s. The second of these regions (region 2) we relate with the ionization of heavy elements. Depending on the model this region ranges from an acoustic depth of $900$ s to $1600$ s, between the region where light elements ionize and the base of the convective zone. Comparing Figures \ref{fig: 4} and \ref{fig: 6} it is possible to notice that the abrupt variation of the gradient in region 2 corresponds in location to the ionization of the last electron of the L-shell.


\section{The footprint of heavy elements ionization in the oscillations}
\label{sec:4}

\begin{figure}
	\includegraphics[width=0.5\textwidth]{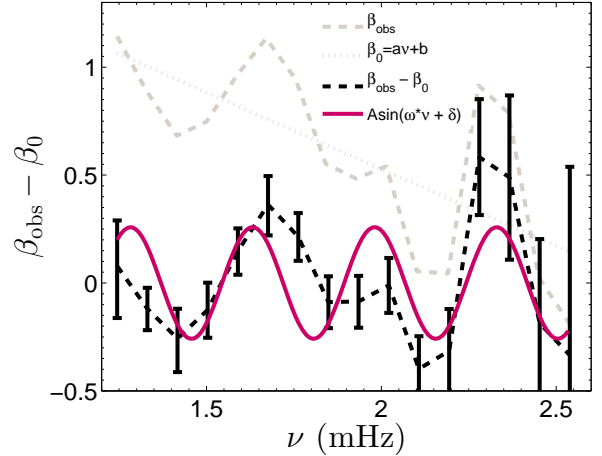}
	\caption{Dashed gray line represents the observational parameter $\beta_{\text{obs}}$ and the dotted gray line is a linear polynomial fit of the type $\beta_0 = a \nu + b$ $(a=-0.7099 \;\; \text{and}\;\; b=1.948)$ to the observational parameter $\beta_{\text{obs}}$. Dark red line shows the sinusoidal, least squares fit, to the difference $\beta_{\text{obs}}-\beta_0$.}
	\label{fig: 7}
\end{figure}

As it is well known, steep variations of the acoustic structure of a star can produce an oscillatory component in the eigenfrequencies of the star \citep{1988IAUS..123..151V, 1990LNP...367..283G, 1994A&A...283..247M} 
\begin{equation}
	\nu_{n,l} \propto \sin (\omega^* \nu_{n,l} + \delta)
\end{equation}
where $\omega^* = 4\pi \tau^*$ and
\begin{equation}
	\tau^* = \int_{r^*}^R \frac{dr}{c}
\end{equation}
is the acoustic depth of the steep variation region. $R$ is the radius of the star, $r^*$ the radial distance to the location of the steep variation region and $c$ represents the adiabatic sound speed. This oscillatory feature can be found not only directly in the eigenfrequencies but also in other observable seismic parameters \citep{1994MNRAS.268..880R, 1997MNRAS.288..572B, 2001A&A...368L...8M, 2004A&A...423.1051B, 2014ApJ...782...16B}. Therefore, the oscillatory feature of the observational $\beta(\nu)$ is intrinsically related to the location where the waves are scattered.
To extract this information from the observable $\beta(\nu)$ we perform as follows. First we do a linear polynomial fit to the observational $\beta(\nu)$, to account, roughly, for the contribution of the superadiabatic zone. Then we subtract this contribution from the observational signature $\beta_{\text{obs}}$. Finally, this difference is fitted to a sinusoid, using a least--squares procedure weighted by error bars values. Figure \ref{fig: 7} shows the difference $\beta_{\text{obs}}-\beta_0$ and the corresponding sinusoidal fit. The resulting value for the angular frequency is $\omega^* = 17.34 \pm 4.52$, meaning a  cyclic frequency of $\nu^* = \omega^*/{2 \pi}$. Then, the period of the oscillation is $T^*=1/\nu^*$, which results in an acoustic depth of $\tau^* \approx 1/{(2T^*)} = 1380$ s. This result shows that contrarily to what happens in the Sun, the oscillatory feature in the observable $\beta(\nu)$ is dominated by a scattering region located between the zone of the second ionization of helium and the base of the convective zone. Taking into account the ionization profiles studied in section \ref{sec:3}, specifically the gradient of the total mean charge, we associate this oscillatory feature with the ionization of heavy elements: carbon, nitrogen and oxygen. 
This location is right between the zone of the second ionization of helium and the base of the convective zone, in the adiabatically stratified part of the convective zone. 
Figures \ref{fig: 4}, \ref{fig: 5} and \ref{fig: 6} show that this is the region where heavy elements, namely CNO, lose their last electrons of the L-shell. From the point of view of the equation of state (EOS), this is an interesting region in the star, because EOS defines the first adiabatic exponent, and this in turn, defines the structure of the outer layers. The treatment and inclusion of processes for a multicomponent mixture of charges with all the complexities associated (e.g. non ideal effects) in the equation of state is not an easy task. We think that part of the mismatches we observe when comparing the theoretical and the observational seismic parameter $\beta(\nu)$ (Figure \ref{fig: 3}) could be due to a poor description of ionization in the equation of state. 
		
We have applied our analysis of the periodicities also to the theoretical $\beta(\nu)$ for all the four models. With this intent we have considered a window of frequencies similar to the observational window and with the same number of points. For all the models the dominant period of the sinusoid is related with the zone of the second ionization of helium which is coherent from the perspective that the equation of state does not include an appropriate description of the ionization processes occurring in the outer layers of the star.
				
The periodicity we observe in the observational $\beta(\nu)$ and which we relate with the gradient of the effective charge $\bar{q}$ is dominated by the ionization of heavy elements. Moreover, \citet{2012A&A...540A..31M} using a different technique based on the second differences, detected an acoustic sharp feature on this star, located around 1370 s. They exclude the possibility of an artefact in the data but they do not attribute physical meaning to the feature. We think it is rather relevant that two distinct methods point to the same scattering location of acoustic modes in the outer layers of this star and we are suggesting in this paper a possible physical meaning to the origin of the unexpected oscillatory feature. \citet{2014ApJ...782...16B} also detected a region of rapid variation of the sound speed in the star $\alpha$ Centauri A, located 6\% under the surface of the star in this deep part of the convection zone.
		
A new generation of equations of state is being developed \citep{2013ASPC..479...11B, 2013CoPP...53..392G} with the intent of providing an accurate description of the first adiabatic exponent in the convection zone. 
For the solar case, the results of a helioseismic inversion for the first adiabatic exponent looks promising \citep{2010Ap&SS.328..147B, 2013MNRAS.430.1636V}. 
This work shows that the utilization of these equations of state to do the modeling of other stars with different chemical compositions and different internal structures, will surely improve our knowledge on stellar structure and evolution.
		
\section{Conclusions}
\label{sec:5}
		
The outer layers of solar-type stars have a direct influence on the frequencies of the acoustic modes. Acoustic glitches associated with partial ionization zones of light elements, particularly, the zone of the second ionization of helium have been detected for several solar-type stars from their oscillatory signatures in the frequencies \citep{2014ApJ...782...18M}. We used a seismic diagnostic, which acts as a proxy of the phase shift of the reflected sound waves, to explore the outer layers of the star HD 49933. Namely, all the region confined between the superadiabatic zone and the base of the convective zone. The method allowed to identify a seismologic relevant oscillatory signature in the frequencies, located between the zone of the second ionization of helium and the base of the convective zone, approximately 5\% under the surface of the star. A detailed study of the ionization profiles of this star lead us to relate this signature with the ionization of heavy elements in the deep convection zone, more precisely, with the ionization of the last electron of the L-shell, in elements such as carbon, nitrogen and oxygen. Further investigation is needed to clarify and understand the effects of ionization processes (and all the complex physics associated) in stellar structure and the subsequently effects in the oscillation spectrum.
In hot solar-type stars, like the star HD 49933, the pattern of partial ionization zones in the interior is expected to be different to that of a cool star like the Sun. The absence of an explicit signature in the phase shift located in the CNO region, for cool stars like the Sun, should not be a surprise since this region is located deeper in these stars, overpassing the base of the convective zone. 

Saha's ionization theory explains with success many features related with atmospheric and spectral characteristics of different stars. However, in the interior the situation seems to be different since the integration with the equation of state is not straightforward. If indeed, the oscillatory signature we detected in the seismic parameter  $\beta(\nu)$ is related to the ionization of heavy elements, it is clear that the current theoretical models underestimate the effects of partial ionization of heavy elements in the outer layers of solar-type stars. Seismology may be starting to reveal that the current versions of the equation of state are not sufficient to describe the properties of the adiabatically stratified part of the convection zone in solar-type stars.
		
Finally and concerning the magnetic activity, it is known that HD 49933 is an active star and a fast rotator. It fits the relation between cycle and rotation periods proposed by \citet{1984ApJ...279..763N} and \citet{2007ApJ...657..486B}. The first stellar cycle variations of the p-mode frequencies were detected precisely in this star by \citet{2010Sci...329.1032G}. The analysis of low-degree frequencies allowed \citet{2006ApJ...640L..95V} to detect solar-cycle variations of the $\Gamma_1$  dip located at the region of the second ionization of helium.  These variations seem to suggest that partial ionization processes in stellar interiors' could have a tight relation with the magnetic activity of the stars. This plausible relation between ionization and magnetism is worth being studied in the future.

\section*{Acknowledgements}

We thank the referee for comments and suggestions that led to a more accurate and robust manuscript. 
We are also grateful to P. Morel for making available the CESAM code for stellar evolution,
to Jordi Casanellas for the modified version of the same code, and to J. Christensen-Dalsgaard for his Aarhus adiabatic pulsation code (ADIPLS). This work was supported by grants from "Funda\c c\~ao para a Ci\^encia e Tecnologia" (SFRH/BD/74463/2010).
%
%

%
%
%

%
%
%
%
%
%

\bsp	
\label{lastpage}
\end{document}